\documentclass[12pt,oneside,nofootinbib,nobibnotes,aps]{article}
\usepackage[cp1251]{inputenc}
\usepackage[english]{babel}
\usepackage{amsmath}
\usepackage{amsfonts}
\usepackage{amssymb}
\usepackage{bm}
\begin{document}
\begin{center}
\textbf{Classical electrodynamics in a space with spin noncommutativity of coordinates}\\\bigskip

V.M.~Vasyuta$^1$, V.M.~Tkachuk$^2$\\
\small\textit{
	E-mail: $^1$waswasiuta@gmail.com, $^2$voltkachuk@gmail.com\\
	Department for Theoretical Physics, Ivan Franko National University of Lviv,\\
	12, Drahomanov St., Lviv, UA-79005, Ukraine}
\end{center}

\begin{quote}\small
	We propose a new relativistic Lorentz-invariant spin-noncommutative algebra. 
	Using the Weyl ordering of noncommutative position operators, we build an analogue of the Moyal-Groenewald product for the proposed algebra. The  Lagrange function of an electromagnetic field in the space with spin noncommutativity is constructed. In such a space electromagnetic field becomes non-abelian. A gauge transformation law of this field is also obtained. Exact nonlinear field equations of noncommutative electromagnetic field are derived from the least action principle. Within the perturbative approach we consider field of a point charge in a constant magnetic field and interaction of two plane waves. An exact solution of a plane wave propagation in a constant magnetic and electric fields is found.\bigskip
	
	
\end{quote}

\section{Introduction}Noncommutativity of position operators appears in string theory \cite{ConnesDouglasSchwarz1998,seiberg1999string} and quantum gravity \cite{DFR}. In \cite{seiberg1999string} it was shown that coordinates on D-brane in a constant Neveu-Schwarz B-field satisfy the following commutation relations
\begin{eqnarray}
\label{CN}
[X_i,X_j]=i\theta_{ij},
\end{eqnarray}
where $\theta_{ij}$ is a constant antisymmetric tensor. This type of noncommutativity is called canonical noncommutativity. The same noncommutativity appears in compactifying the IKKT M-theory \cite{ConnesDouglasSchwarz1998}. In quantum gravity the noncommutativity can be thought of as a phenomenological effect from quantum space-time, which incorporates the notion of the minimal length into ordinary physics. Moreover, noncommutativity arises even in the pure classical mechanics. It can be shown that coordinates of a charged particle with a small mass in a strong magnetic field do not commute (the corresponding Poisson bracket is non-zero) \cite{Pei, Jac}.

It is interesting that, decades before string theory, in searching for a generalization of the ordinary commutation relations between the operators of dynamical variables Snyder developed a noncommutative Lorentz-invariant algebra of the following form
\begin{eqnarray}
\label{Snaid}
[X_\mu,X_\nu]=il^2L_{\mu\nu},
\end{eqnarray} 
where $L_{\mu\nu}$ are generators of the Lorentz rotation group, $l$ is a small parameter \cite{Sn,Sn1}.

One of the biggest problem of canonical noncommutativity (\ref{CN}) with a constant right-hand part of coordinate commutator is a violation of a rotational symmetry and as a consequence existing of a chosen direction.

Rotational invariance (or more widely Lorentz-invariance) can be restored in noncommutative spaces by replacing constants $\theta_{ij}$ with some more complicated objects. There are already several rotational (Lorentz) invariant noncommutative algebras. First of all, the above-mentioned Snyder algebra is Lorentz-invariant (\ref{Snaid}). The other example of such an algebra can be built by assuming that $\theta_{ij}$ are operators commuting with each other and transforming as components of tensor \cite{DFR}. Objects $\theta_{ij}$ can be composed from some additional degrees of freedom. In this way by using coordinates of an additional harmonic oscillator rotational invariant noncommutativity was introduced in \cite{Gnat,krynytskyi}.

Recently, there were proposed several rotational invariant noncommutativities, where coordinate commutators are postulated to be proportional to some functions of spin operators. These algebras are known in literature as spin noncommutativity or noncommutativity due to spin.

In \cite{fg} the following algebra with spin noncommutativity was proposed
\begin{eqnarray}
\label{GG}
	\begin{array}{c}
	\displaystyle\left[X_i,X_j\right]=i\hbar\theta^2\varepsilon_{ijk}s_k,\quad\left[X_i,P_j\right]=i\hbar\delta_{ij},\quad\left[P_i,P_j\right]=0,\\[6pt]
	\displaystyle\left[s_i,s_j\right]=i\hbar\varepsilon_{ijk} s_k,\quad\left[X_i,s_j\right]=i\hbar\theta\varepsilon_{ijk}s_k,
	\end{array}
\end{eqnarray}
where $\theta$ is the parameter of spin noncommutativity, $s_k$ is the $k$-th component of spin operator, $\delta_{ij}$ is the Kronecker delta-symbol, $\varepsilon^{ijk}$ is the Levi-Civita tensor. This algebra is obtained by shifting position operators by spin operators.
\begin{eqnarray}
\label{pred}
X_i=x_i+\theta s_i,\qquad P_i=p_i.
\end{eqnarray}
It is easy to see that algebra (\ref{GG}) is invariant with respect to spatial rotations. 

In \cite{fg} supersymmetric extension of a harmonic oscillator was considered in a space with spin noncommutativity (\ref{GG}). It was shown the degeneracy of the ground state of this oscillator with spontaneously broken symmetry with respect to rotations. Using obtained results a new phase was predicted in systems with dominating dipole-dipole interactions (e.g. bose-condensate of $^{52}$Cr).

Topology of the Aharonov-Bohm effect in space (\ref{GG}) was investigated in \cite{fg1}. It was shown that the leading contribution due to the noncommutativity is inherited from commutative problem (i.e. topological properties does not deform, although the field has strong anisotropy at small scales $ r\sim\theta $).

Using this algebra (with replacing  $ \theta\rightarrow i\theta $) it is possible to explain the triplet Cooper pairing mechanism, which was found in some exotic superconductors: for any non-zero $\theta$ there exist triplet states with energy smaller than for singlet states \cite{fg2}. 

In \cite{Vasyuta2} corrections to the hydrogen atom energy spectrum due to spin noncommutativity (\ref{GG}) were found. Logarithmic dependence of $s$-levels energy on $\theta$ was obtained within the modified perturbation theory, which was developed there. Moreover, an upper bound for parameter of noncommutativity was estimated using a precise measurement of the transition frequency between $2s-1s$ energy levels.

The other spin noncommutativity \cite{GKdS} was built by introducing the following position operators $X^\mu=x^\mu+\theta W^\mu$, where $W^\mu=\frac{1}{2}\varepsilon^{\mu\nu\rho\sigma}S_{\nu\rho}p_\sigma$ is the Pauli-Lubansky pseudo-vector, $S_{\nu\rho}=\frac{i}{4}[\gamma_\nu,\gamma_\rho]$. The corresponding algebra reads
\begin{eqnarray}
	\label{GkS}
\begin{array}{c}
\displaystyle\left[X^\mu,X^\nu\right]=i\theta\hbar\varepsilon^{\mu\nu\rho\sigma}S_{\rho\sigma}-i\theta^2\varepsilon^{\mu\nu\rho\sigma}W_\rho P_\sigma,\\[6pt]
\displaystyle\left[X^\mu,P^\nu\right]=-i\hbar\eta^{\mu\nu},\quad \left[P^\mu,P^\nu\right]=0,\quad\left[X^\mu,\gamma^\nu\right]=-\frac{i\theta}{2}\varepsilon^{\mu\nu\rho\sigma}P_\rho\gamma_\sigma,
\end{array}
\end{eqnarray}
where $\eta^{\mu\nu}$ is the Minkowski tensor. Such a coordinate noncommutativity appears in quantum twistors theory \cite{LW} and within the quasiclassical approximation of a motion of a spinning particle in curved space-time \cite{ramirez2014frenkel,ramirez2015frenkel}. One of the main advantages of this algebra is its relativistic invariance. In space with this algebra the Dirac equation was studied in \cite{FGK}. The conservation law for an electrical current was estimated. In addition, there it was shown that noncommutativity breaks the degeneracy of energy levels for an electron in a constant magnetic field. Although (\ref{GkS}) is Lorentz invariant, it breaks micro-causality \cite{HW}.

A non-relativistic reduction of (\ref{GkS}) was built by shifting the commutative position operators by the 3D analogue of the Pauli-Lubansky vector $ W_i=\frac{1}{4}\varepsilon_{ijk}\sigma_jp_k $ \cite{Vasyuta}. The obtained algebra reads
\begin{eqnarray}
\label{neralg}
\begin{array}{c}
\displaystyle
\left[X_i,X_j\right]=i\theta\varepsilon_{ijk}s_k+i\frac{\theta^2}{4\hbar}\varepsilon_{ijk}P_k(\textbf{s},\textbf{P}),\quad
	\left[X_i,P_j\right]=i\hbar\delta_{ij},\quad\left[P_i,P_j\right]=0,\\
\displaystyle
\left[s_i,s_j\right]=i\hbar\varepsilon_{ijk}s_k,\quad
\left[X_i,s_j\right]=i\frac{\theta}{2}(P_js_i-\delta_{ij}(\textbf{s},\textbf{P})).
\end{array}
\end{eqnarray}
In space with this algebra a minimal length is present. An exact solution of harmonic oscillator in this spin noncommutativity (\ref{neralg}) was found \cite{Vasyuta}.

Electrodynamics in space with canonical noncommutativity is well studied on both classical and quantum levels \cite{douglas2001noncommutative, szabo2003quantum, berrino2003noncommutative, chaichian2001hydrogen, buric2002one,guralnik2001testing, cai2001superluminal, castorina2004noncommutative, adorno2011classical, gaete2004coulomb,castorina2005vacuum, banerjee2004noncommutative, carlson2002noncommutative,madore2000gauge}. As far as we know, in space with spin noncommutativity the electromagnetic field has not been considered before.

In the present paper we build a new Lorentz-invariant coordinate noncommutativity. In space with proposed algebra we define an analogue of the Moyal product. An unique feature of the obtained product is the decomposition of this product into a product of two matrix function objects, which can be associated with a function in space with spin noncommutativity. In space with this algebra we develop a classical electrodynamics. Within the noncommutative electrodynamics we consider a field of a point charged particle in a constant magnetic field, interaction of two plane waves, propagation of a plane wave in a constant magnetic and electric field (an exact solution).

\section{Relativistic Lorentz-invariant spin noncommutativity}

Direct relativistic extension of the algebra with spin noncommutativity (\ref{GG}) can be built by replacing the Pauli matrices (which generate the Clifford algebra $\mathcal{C}\ell(3)$) with the Dirac gamma matrices ($\mathcal{C}\ell(1,3)$)  in representation (\ref{pred}). Obtained noncommutative position operators become matrices and are equal to ordinary coordinates shifted by the corresponding matrices $\gamma^\mu$
\begin{eqnarray}
\label{SNR_repr}
X^\mu=x^\mu+i\theta\gamma^\mu,
\end{eqnarray}
where the imaginary unit is added for the hermiticity of space coordinates. By assuming that momenta in noncommutative space are the same as in the commutative case $P^\mu=p^\mu=i\partial^\mu$, it is easy to write down the full closed algebra
\begin{eqnarray}
\label{alg}
\begin{array}{c}
\left[X^\mu,X^\nu\right]=2i\theta^2\sigma^{\mu\nu},
		\quad\left[X^\mu,\sigma^{\alpha\beta}\right]=2\theta(\gamma^\alpha\eta^{\mu\beta}-\gamma^\beta\eta^{\mu\alpha}),\\
\left[X^\mu,P^\nu\right]=-i\eta^{\mu\nu},\quad\left[P^\mu,P^\nu\right]=0,
		\quad\left[P^\mu,\sigma^{\alpha\beta}\right]=0,\\
\left[\sigma_{\alpha\beta},\sigma_{\gamma\delta}\right]=i\left(\eta_{\alpha\gamma}\sigma_{\beta\delta}-\eta_{\beta\gamma}\sigma_{\alpha\delta}-
		\eta_{\alpha\delta}\sigma_{\beta\gamma}+\eta_{\beta\delta}\sigma_{\alpha\gamma}\right),
\end{array}
\end{eqnarray}
where $\sigma^{\mu\nu}=i\left[\gamma^\mu,\gamma^\nu \right]/2$. Note that both $\sigma^{\mu\nu}$ and $L^{\mu\nu}$ are generators of $SO(1,3)$ but in different representations. In such a context coordinate noncommutativity (\ref{alg}) is similar to the Snyder one (\ref{Snaid}).

To develop any theory in a noncommutative space the first thing one must do is to find a rule how to construct a noncommutative counterpart for a given commutative function $f(x)$. With the necessity, a problem of the position operators ordering appears on such a transition to noncommutative space. This problem is traditionally solved by introducing the Moyal-Groenewald star product, where the Weyl ordering of operators is used
\begin{eqnarray}
\label{Moyal}
	(f\star g)(X)=\int d^nk\int d^nqe^{ik_\mu X^\mu}F_ke^{iq_\mu X^\mu}G_q,
\end{eqnarray}
where $F_k=1/(2\pi)^n\int d^nxe^{ikx}f(x)$, $G_k=1/(2\pi)^n\int d^nxe^{ikx}g(x)$ are the Fourier components of $f(x)$ and $g(x)$ respectively. In the case of canonical noncommutativity the Moyal (Moyal-Groenewald) product is expressed in terms of commutative functions in the following manner $f\star g=f(x)\exp({i}/{2}\overleftarrow{\partial}_i\theta^{ij}\overrightarrow{\partial}_j)g(x)$. 

The star product with qualitatively different properties appears in a space with coordinates (\ref{SNR_repr}). In this space the variables of integration in (\ref{Moyal}) can be separated
	\begin{eqnarray}
	\label{P1}
	\int d^4k\int d^4qe^{i(kX)}F_ke^{i(qX)}G_q=\widehat{T}\int d^4ke^{i(kx)}F_k\cdot \widehat{T}\int d^4qe^{i(qx)}G_q=\tilde{f}(x)\tilde{g}(x), 
	\end{eqnarray}
where $\tilde{f}(x)=\widehat{T}f(x)$, $\widehat{T}=e^{i\theta(\gamma\partial)}$, $(AB)=\eta^{\mu\nu}A_\mu B_\nu$ denotes a (3+1)D scalar product. It is easy to see that we can identify the matrix function $\tilde{f}(x)=\widehat{T}f(x)$ as a noncommutative equivalent of commutative $f(x)$. It is remarkable that noncommutative function is equal to the action of the translation operator $\widehat{T}=e^{i\theta(\gamma\partial)}$ on the commutative function. Decomposition of the variables of integration in (\ref{P1}) leads to algebraic properties of this product, which differ from properties of the Moyal product in canonical noncommutativity. It can be already seen from the definitions of a single function in these spaces. In canonical noncommutativity a single function can be defined as the Moyal product of this function with a unit element $f(x)\star1=f(x)$ and is equal to ordinary commutative function. So, on the contrary to the canonical noncommutativity the spin noncommutativity influences a single function $\tilde{f}(x)=\widehat{T}f(x)$.

Let us analyze the ordering of gamma matrices in the expression $\widehat{T}f(x)$. Action of $T=e^{i\theta(\gamma\partial)}$ on $f(x)$ is equal to the Taylor expansion
\begin{eqnarray}
\label{exp}
	\tilde{f}=e^{i\theta(\gamma\partial)}f(x)=\sum\limits_{n=0}^\infty\frac{i^n\theta^n}{n!}(\gamma\partial)^nf(x)=\sum\limits_{n=0}^\infty\frac{(i\theta)^n}{n!}f^{(n)}_{\mu_1\mu_2...\mu_n}\gamma^{\mu_1}\gamma^{\mu_2}\cdot...\cdot\gamma^{\mu_n},
\end{eqnarray}
where $(n)$ denotes the order of derivative of $f(x)$. Inasmuch as derivatives commute, multiplier with fixed number of derivatives with respect to every coordinate is equal to symmetrized product of gamma matrices. This expression is equal to the Taylor expansion of $f(x+i\theta\gamma)$ in $\theta$, where every term consist of all symmetrized permutations of gamma matrices. For instance, in the case of $n=3$, $\mu_1=x,\;\mu_2=x,\;\mu_3=y$ the corresponding term in expansion reads
\begin{eqnarray*}
\frac{(i\theta)^3}{3!}f^{'''}_{xxy}\cdot(2\gamma_x\gamma_x\gamma_y+2\gamma_x\gamma_y\gamma_x+2\gamma_y\gamma_x\gamma_x).
\end{eqnarray*}

On the other hand, using the identity $(\gamma\partial)^2=-\partial^2$ one can rewrite $\widehat{T}$ in the following form
\begin{eqnarray}
\label{TCS}
	\widehat{T}=e^{i\theta(\gamma\partial)}=\tilde{C}+i\theta(\gamma\tilde{\partial}),
\end{eqnarray}
where $\tilde{C}=\cos(\theta\sqrt{-\partial^2})$,  $\tilde{\partial}_\mu=\text{sinc}(\theta\sqrt{-\partial^2})\partial_\mu$, $\text{sinc}(x)=\sin(x)/x$. This representation is useful in deriving some properties of noncommutative product. Firstly, using the Taylor expansion of $\widehat{T}$ it can be showed that the following natural properties are satisfied $\tilde{1}=1$,  $\widehat{T}x^\mu=x^\mu+i\theta\gamma^\mu=X^\mu$. Expression for a product of two noncommutative function $\tilde{f}$ and $\tilde{g}$ can be obtained in the following form
\begin{eqnarray}
\left[\tilde{C}+i\theta(\gamma\tilde{\partial})\right]f\cdot\left[\tilde{C}+i\theta(\gamma\tilde{\partial})\right]g=\tilde{C}f\tilde{C}g-\theta^2(\gamma\tilde{\partial})f(\gamma\tilde{\partial})g+i\theta\left((\gamma\tilde{\partial})f\tilde{C}g+\tilde{C}f(\gamma\tilde{\partial})g\right).
\end{eqnarray}
Using the well-known identity $(\gamma A)(\gamma B)=(AB)-i\sigma^{\mu\nu}A_\mu B_\nu$ and identities $\tilde{C}f\tilde{C}g-\tilde{\partial}^\mu f\tilde{\partial}_\mu g=\tilde{C}(fg)$ and $\tilde{\partial}_\mu f\tilde{C}g+\tilde{C}f\tilde{\partial}_\mu g=\tilde{\partial}_\mu(fg)$, which can be obtained from the definition, one can find
\begin{eqnarray}
\label{prodd}
	\widehat{T}(f)\widehat{T}(g)=\widehat{T}(fg)+i\theta^2\sigma^{\mu\nu}\tilde{\partial}_\mu f\tilde{\partial}_\nu g.
\end{eqnarray}
This expression shows the connection between product of two noncommutative functions $\widehat{T}(f)\widehat{T}(g)$ and a noncommutative counterpart of the commutative product of these functions. It can be showed by direct calculations that such a product of noncommutative functions is associative
\begin{eqnarray}
	\tilde{f}(\tilde{g}\tilde{h})=	(\tilde{f}\tilde{g})\tilde{h}.
\end{eqnarray}

From (\ref{prodd}) it follows that $(\tilde{f})^n=\widehat{T}(f^n)$. Action of $\widehat{T}$ on a composite function $\varphi(f(x))$ can be obtained in the same way. Acting by $\widehat{T}$ on the Taylor expansion of $\varphi(f(x))$ and using (\ref{prodd}) one obtain
\begin{eqnarray}
\label{comp_func}
\widehat{T}{\varphi(f)}=\varphi(\widehat{T}f).
\end{eqnarray}

Since $\widehat{T}$ commutes with derivative, the differentiation of a noncommutative function can be easily performed $\partial^\mu\widehat{T}(f(x))=\widehat{T}(\partial^\mu f(x))$. The differentiation of a composite function is a bit more complicated
\begin{eqnarray}
\begin{array}{c}
\partial_\rho\tilde{\varphi}(f)=\widehat{T}\left(\frac{\partial f}{\partial \varphi}\partial_\rho\varphi\right)
=\widehat{T}\left(\frac{\partial f}{\partial \varphi}\right)
\widehat{T}\left(\partial_\rho\varphi\right)-i\theta^2\sigma^{\mu\nu}\left(\partial_\mu\frac{\partial f}{\partial \varphi}\right)\left(\partial_\nu\partial_\rho\varphi\right)\;\text{or}
\\
\partial_\rho\tilde{\varphi}(f)=\widehat{T}\left(\partial_\rho\varphi\frac{\partial f}{\partial \varphi}\right)
=\widehat{T}\left(\partial_\rho\varphi\right)\widehat{T}\left(\frac{\partial f}{\partial \varphi}\right)
+i\theta^2\sigma^{\mu\nu}\left(\partial_\mu\frac{\partial f}{\partial \varphi}\right)\left(\partial_\nu\partial_\rho\varphi\right).
\end{array}
\end{eqnarray}

In the next section we will build an action of a noncommutative electromagnetic field. The action should be a scalar. Therefore let us define the integration of a noncommutative function $\tilde{f}$ as $(1/4)\int dx\;\text{Sp}\{\tilde{f}\}$. Inasmuch as gamma matrices are traceless, we have $(1/4)\int dx\;\text{Sp}\{\tilde{f}\}=\int dx f(x)$. Since $\sigma^{\mu\nu}$ are also traceless, integral of product of two function in noncommutative space is the same as in the commutative case $(1/4)\int dx\;\text{Sp}\{\tilde{f}\tilde{g}\}=\int dx fg$. In such a manner it can be showed that integral of a product of three noncommutative functions do not feel the noncommutativity too. And only integral of a product of four noncommutative functions changes due to noncommutativity. Using the equality $\text{Sp}\{\sigma^{\mu\nu}\sigma^{\alpha\beta}\}=g^{\mu\alpha}g^{\nu\beta}-g^{\mu\beta}g^{\nu\alpha}$, it can be showed that $\int dx\;\text{Sp}\{\tilde{f}\tilde{g}\tilde{h}\tilde{j}\}\neq\int dx fghj$.

\section{Electromagnetic field in a space with spin noncommutativity}
\subsection{Tensor of electromagnetic field. Action for electromagnetic field}

Potential of an electromagnetic field in noncommutative space should be associated with the matrix function $\tilde{A^\mu}=\widehat{T}A^\mu$ in the manner described in Section 2. Noncommutative electromagnetic field $\tilde{A^\mu}$ is non-abelian, so commutator of different components of the potential equals
	\begin{eqnarray}
	\left[e^{i\theta(\gamma\partial)}A^\mu,e^{i\theta(\gamma\partial)}A^\nu\right]=2i\theta^2\sigma^{\alpha\beta}\tilde{\partial}_\alpha A^\mu\tilde{\partial}_\beta A^\nu.
	\end{eqnarray}
Then tensor of such an electromagnetic field reads
	\begin{eqnarray}
	\tilde{F}^{\mu\nu}=\partial^\mu\tilde{A}^\nu-\partial^\nu\tilde{A}^\mu-ie\left[\tilde{A}^\mu,\tilde{A}^\nu\right]=
	\partial^\mu \tilde{A}^\nu-\partial^\nu \tilde{A}^\mu+2e\theta^2\sigma^{\alpha\beta}\tilde{\partial}_\alpha A^\mu\tilde{\partial}_\beta A^\nu.
	\end{eqnarray}

The Lagrange function and the action can be found in the following form
\begin{eqnarray}
\label{act}
S=\frac{1}{4}\int d^4x\,\text{Sp}\left\{-\frac{1}{4}\tilde{F}^{\mu\nu}\tilde{F}_{\mu\nu}\right\}.
\end{eqnarray}
After tracing it can be rewritten as follows
\begin{eqnarray}
\label{act_2}
S=\int d^4x\left\{-\frac{1}{4}{F}^{\mu\nu}{F}_{\mu\nu}
-\frac{1}{4}e^2\theta^4\left(
\tilde{\partial}^\alpha A^\mu\tilde{\partial}_\alpha A_\mu\tilde{\partial}^\beta A^\nu\tilde{\partial}_\beta A_\nu-\tilde{\partial}^\alpha A^\mu\tilde{\partial}_\alpha A_\nu\tilde{\partial}^\beta A^\nu\tilde{\partial}_\beta A_\mu
\right)\right\}.
\end{eqnarray}
It is evident that the action (\ref{act_2}) is invariant with respect to Lorentz transformations as it is written implicitly in a covariant form. Also it is easy to see that it is $\mathcal{C}$-, $\mathcal{P}$- and $\mathcal{T}$- invariant.
From (\ref{act_2}) it follows that corrections due to the noncommutativity are of the fourth order in $\theta$.

\subsection{Gauge transformations}
Gauge transformation for the electromagnetic field in noncommutative space differs from ordinary $U(1)$ transformations. We find the gauge transformation law for the noncommutative electromagnetic field $\tilde{A}^\mu$ from the condition of a gauge invariance of a covariant derivative under the multiplication by $\exp{(ie\tilde{\alpha})}$
\begin{eqnarray}
	\mathcal{D}'_\mu(e^{ie\tilde{\alpha}}\tilde{\varphi})=\partial_\mu(e^{ie\tilde{\alpha}}\tilde{\varphi})+ie\tilde{A}'_\mu e^{ie\tilde{\alpha}}\tilde{\varphi}=e^{ie\tilde{\alpha}}(\partial_\mu\tilde{\varphi}+ie\tilde{A}_\mu\tilde{\varphi})=e^{ie\tilde{\alpha}}\mathcal{D}_\mu\tilde{\varphi},
\end{eqnarray}
where $\mathcal{D}=\partial+ieA$, $\varphi$ is the matter field, the prime denotes a variable after the transformation. 

Differentiating by parts the first term in the left-hand side, we obtain $$ie\tilde{A}'_\mu e^{ie\tilde{\alpha}}\tilde{\varphi}=e^{ie\tilde{\alpha}}ie\tilde{A}_\mu e^{-ie\tilde{\alpha}}(e^{ie\tilde{\alpha}}\tilde{\varphi})-(\partial_\mu e^{ie\tilde{\alpha}})e^{-ie\tilde{\alpha}}(e^{ie\tilde{\alpha}}\tilde{\varphi}).$$ Taking into account the arbitrariness of $\tilde{\varphi}$ we find the following gauge transformation law
\begin{eqnarray}
	\tilde{A}'_\mu =e^{ie\tilde{\alpha}}\tilde{A}_\mu e^{-ie\tilde{\alpha}}+\frac{i}{e}\cdot e^{ie\tilde{\alpha}}(\partial_\mu e^{-ie\tilde{\alpha}}).
\end{eqnarray}
The shape of this transformation law is similar to the gauge transformation law for a non-abelian field in a commutative space. After commuting $A_\mu$ and $e^{-ie\tilde{\alpha}}$ and differentiating $e^{-ie\tilde{\alpha}}$ we obtain
\begin{eqnarray}
\label{gg}
\tilde{A'}_\mu=\tilde{A}_\mu+\partial_\mu\tilde{\alpha}-i\theta^2\sigma^{\nu\beta} (e^{ie\tilde{\alpha}}\tilde{\partial}_\nu e^{-ie\alpha})\tilde{\partial}_\beta(2A_\mu+\partial_\mu\alpha).
\end{eqnarray}
The corresponding infinitesimal transformations read
\begin{eqnarray}
\label{gauge}
\tilde{A'}_\mu=\tilde{A}_\mu+\partial_\mu\tilde{\alpha}-2e\theta^2\sigma^{\nu\beta}\tilde{\partial}_\nu{\alpha}\tilde{\partial}_\beta A_\mu.
\end{eqnarray}
As in the case of the Yang-Mills field it is easy to find a gauge transformation law for $\tilde{F}^{\mu\nu}$ and show action's invariance with respect to the gauge transformations. The gauge transformations for $A_\mu$ are more complicated and can be obtained by acting $\widehat{T}^{-1}=e^{-i\theta(\gamma\partial)}$ on the (\ref{gg}).

Algebra of the gauge group is $u(1)\times so\gamma(1,3)$, where $so\gamma(1,3)$ consist of $\gamma^\mu$, $\sigma^{\mu\nu}$ and the unit matrix. The corresponding gauge group is $U(1) \times SL(1,3)$. The first multiplier reflects intrinsic symmetry of an electromagnetic field and the latter one is present due to the noncommutative structure of a space-time.

Moreover, an algebra of a gauge group of an arbitrary non-abelian field in noncommutative space can be written as a direct product $g\times so\gamma(1,3)$, where $g$ is the algebra of the gauge group $G$ of the field in commutative space. But in this case because of the non-abelian character of $G$ the noncommutative gauge group is not simply a direct product $G\times SL(1,3)$.

\subsection{Field equations}
Let us find equations of the electromagnetic field in noncommutative space from the least action principle. After varying the action (\ref{act_2}) we obtain the exact equations
\begin{eqnarray}
\label{eqel}
\partial^2A^\mu+e^2\theta^4\tilde{\partial}_\alpha\left(
\tilde{\partial}^\alpha A^\mu\tilde{\partial}^\beta A^\nu\tilde{\partial}_\beta A_\nu-\tilde{\partial}^\beta A^\mu\tilde{\partial}^\alpha A^\nu\tilde{\partial}_\beta A_\nu
\right)=0.
\end{eqnarray}
It is easy to see that in the limit of commutative space $\theta\rightarrow0$ these equations become the Maxwell equations.

Some general remarks about the structure of these equations can be said. Despite they are nonlinear,  it is easy to see that the equation for a chosen component $A^\mu$ is linear if other components are known. This feature may be helpful in perturbative analysis of these equations. If a certain component $A^\mu$ is constant in commutative space, it satisfies the corresponding equation (\ref{eqel}) automatically. So in this case $A^\mu$ in noncommutative space is the same as in the commutative space. In particular, if $A^\mu=0$ in commutative space, it equals zero in noncommutative space too. 

On the other hand, if there is only one non-zero component $A^\kappa$, the equation for it reads
\begin{eqnarray}
\partial^2A^\kappa+e^2\theta^4\tilde{\partial}_\alpha\left(
\tilde{\partial}^\alpha A^\kappa\tilde{\partial}^\beta A^\kappa\tilde{\partial}_\beta A_\kappa-\tilde{\partial}^\beta A^\kappa\tilde{\partial}^\alpha A^\kappa\tilde{\partial}_\beta A_\kappa
\right)=\partial^2A^\kappa=0
\end{eqnarray}
and coincides with the commutative Maxwell equation. So, noncommutativity does not affect the systems, which are described by vector potential with one non-zero component. Moreover, if a vector potential by the appropriate Lorentz transformation can be reduced to the vector with only one non-zero component, it also does not feel the noncommutativity. Such systems usually are simple and consist of one source of the field, e.g. a point charge, a linear wire with a current, a plane wave, etc. Spin noncommutativity does not influence these systems.

Interaction of the field with a current $j^\mu$ can be easily included by adding the corresponding term in the action (\ref{act_2})  $S_{int}=(1/4)\int dx\,{\rm Sp}(\tilde{j}\tilde{A})=\int dx (jA)$. So, spin noncommutativity does not affect the interaction between the field and a given current. The equations taking into account the sources read
\begin{eqnarray}
\label{eqel_int}
\partial^2A^\mu+e^2\theta^4\tilde{\partial}_\alpha\left(
\tilde{\partial}^\alpha A^\mu\tilde{\partial}^\beta A^\nu\tilde{\partial}_\beta A_\nu-\tilde{\partial}^\beta A^\mu\tilde{\partial}^\alpha A^\nu\tilde{\partial}_\beta A_\nu
\right)=j^\mu.
\end{eqnarray}
reducing to (\ref{eqel}) for $j^\mu=0$.

\section{Electromagnetic field of several systems}
\subsection{Electrostatic field of a point charge in a constant magnetic field}

As it was shown in Section 3, spin noncommutativity (\ref{alg}) does not modify the electrostatic field of a point particle. The situation changes when a particle is placed in a magnetic field. Since electromagnetic field is a non-abelian in the noncommutative space, electric and magnetic fields interact with each other. In particular, such an interaction leads to the modification of the Coulomb law in an external magnetic field.

Influence of a constant magnetic field on the electric field of a charged particle in the space with canonical noncommutativity was obtained by deriving the corresponding corrections to the commutative potential from the Lagrange function of the electromagnetic field \cite{berrino2003noncommutative, gaete2004coulomb, stern2008noncommutative, adorno2011classical}. Up to the first order in $\theta^{ij}$ Coulomb potential reads \cite{adorno2011classical}
\begin{eqnarray}
	\label{adorno}
	A^0(\bm{x})=\frac{q}{r}\left(1-e\left[\frac{1}{r^2}(\bm{x}\bm{B})(\bm{x}\bm{\theta})+(\bm{B}\bm{\theta})\right]\right),
\end{eqnarray}
where $\theta^i=\varepsilon^{ijk}\theta_{jk}$, $r^2=(\bm{x}\bm{x})$, $\bm{B}$ is the magnetic field.

For the sake of completeness let us mention the other approach to study noncommutative Coulomb potential. Within this scheme noncommutative Hamiltonian of hydrogen atom is built by replacing position operators $x$ with their noncommutative counterparts $X$. In this case potential energy becomes an operator. The corresponding Schr\"{o}dinger equation describes a motion of an electron in a modified Coulomb potential. In this way, Hydrogen atom was studied in spaces with canonical noncommutativity \cite{chaichian2001hydrogen, gaete2004coulomb} and spin noncommutativity (\ref{GG}) \cite{Vasyuta2}. 

In this subsection we consider the modification of the Coulomb law in a constant magnetic field due to spin noncommutativity. Let us find the solution of field equations (\ref{eqel_int}) with current $j^\mu=\left(-q\delta(\bm{x});0;0;0\right)^T$ in the following form
\begin{eqnarray}
\label{pr1a}
A^0=\frac{q}{4\pi r}+e^2\theta^4A^0_{(1)}+\mathcal{O}(\theta^8),\quad A^i=\varepsilon^{ijk}h_jx_k+e^2\theta^4A^i_{(1)}+\mathcal{O}(\theta^8),
\end{eqnarray}
where index $(1)$ denotes the first order term in expansion of $A^\mu$ in $\theta^4$.
In the commutative limit $\theta\rightarrow0$ vector potential $A^\mu$ reproduces the electrostatic field of a point charge $-q$ and a constant magnetic field $\bm{h}$.
Substitution of ansatz (\ref{pr1a}) in field equations (\ref{eqel}) gives the following equations for $A^\mu_{(1)}$
\begin{eqnarray}
\nabla^2 A^0_{(1)}=\frac{qh^2}{4\pi r^3}-\frac{3q(\bm{h}\bm{x})^2}{4\pi r^5}+qh^2\delta(\bm{x}),\\
\nabla^2 A^i_{(1)}=\frac{2q^2}{(4\pi)^2r^6}\varepsilon^{ijk}h_jx_k,
\end{eqnarray}
where $\nabla=\bm{e}_x\partial_x+\bm{e}_y\partial_y+\bm{e}_z\partial_z$ is the del operator.

These equations can be easily solved. Finally, the solutions (\ref{pr1a}) read
\begin{eqnarray}
\label{sol1a}
A^0=\frac{q}{4\pi r}\left[1-e^2\theta^4\left({h^2}-\frac{(\bm{h},\bm{e}_r)^2}{2}\right)\right]+\mathcal{O}(\theta^8),\\ 
\label{sol1b}
A^i=\varepsilon^{ijk}h_jx_k\left[1+\frac{e^2\theta^4}{2}\frac{q^2}{(4\pi)^2r^4}\right]+\mathcal{O}(\theta^8),
\end{eqnarray}
where $\bm{e}_r=\bm{x}/r$. As it can be seen from (\ref{sol1a}), the magnetic field shields the charge making its value smaller. In addition, this screening is anisotropic, the biggest effect appears in the perpendicular to $\bm{h}$ directions. In both spin and canonical noncommutativities this effect is distance-independent. But on the contrary to the case of canonical noncommutativity, in spin noncommutativity the shielding never turns zero in any directions and always decreases the charge value. 

On the other hand, it follows from (\ref{sol1b}) that there is an inverse effect: the charge also affects the magnetic field. This influence does not depend on the sign of a charge, but only on its absolute value. This influence is short-range and always effectively increases the value of the magnetic field strength.

\subsection{Interaction of two plane waves}

Since field equations (\ref{eqel}) are nonlinear, the superposition principle is not obeyed, and in general case a sum of two solutions (e.g. plane waves) will not be a solution of these equations. In perturbative analysis it can be treated as interaction of these solutions (in our case plane waves) with each other. In this subsection we consider interaction of two plane waves in space with spin noncommutativity. 

Such a problem for canonical noncommutativity was studied in \cite{berrino2003noncommutative}. The authors within the perturbative scheme have found a corrections to the potential of two waves. As a zero approximation they have considered potential of two waves in the form of a superposition of two complex exponents $Ae^{ikx}$ and $Be^{ik'x}$. But such a trick with considering the complex exponent against of real trigonometric functions cannot be used for nonlinear equations, since one requires from electromagnetic field to be real.

Let us find the solution $A^\mu$ of (\ref{eqel}), which in commutative limit $\theta\rightarrow0$ reduces to $A^\mu\big|_{\theta=0}=A^\mu_{(0)}$, where
\begin{eqnarray}
A_{(0)}^0=0;\;A_{(0)}^1=0;\;A_{(0)}^2=B\sin\Theta\cos(kx);\\
A_{(0)}^3=C\cos(qx)+B\cos{\Theta}\cos(kx).
\end{eqnarray}
In commutative space this potential describes a proliferation of two plane waves with amplitudes $B,\;C$ and wave vectors $k^\mu$ and $q^\mu$. The angle between directions of oscillation of the waves is $\Theta$. %The angle between directions of proliferation of these waves is .

Substitution of an ansatz $A^\mu=A^\mu_{(0)}+A^\mu_{s}$ in (\ref{eqel}) leads to the equation like $\partial^2f=B\cos(kx)$. If $k^\mu$ is not a null vector $k^2\neq0$, the solution is simple and reads $f=-B\cos(kx)/k^2$. But such a solution cannot be used in the case of photonic wave vectors. It can be easily checked that in this case the solution of such an equation can be found in the form $$f=\frac{(ax)}{2(ak)}\sin(kx),$$ where $a^\mu$ is an arbitrary vector satisfying conditions $(ax)\neq0,\;(ak)\neq0$. But this solution describes a resonant increasing of the field and so the energy of the field. It is clear that such a resonant behaviour cannot take place in our system and is an artifact of the performed ansatz.

This problem can be solved within the idea of the so-called Bogolyubov-Krylov method \cite{krylov1949}. According to this scheme the solution of the corresponding equations should be found in the following form
\begin{eqnarray}
\label{s3a}
A^0=0;\;A^1=0\;A^2=B\sin\Theta\cos(kx+e^2\theta^4f_1)+e^2\theta^4a_1,\\
\label{s3b}
A^3=C\cos(qx+e^2\theta^4f_2)+B\cos{\Theta}\cos(kx+e^2\theta^4f_3)+e^2\theta^4a_2.
\end{eqnarray}

Substitution of (\ref{s3a}) and (\ref{s3b}) in the field equations (\ref{eqel}) gives the following equations for auxiliary functions $f_1$, $f_2$, $f_3$, $a_1$, $a_2$
\begin{eqnarray}
\sin(kx)\partial^2f_1-2\cos(kx)k_\alpha\partial^\alpha f_1=
-\frac{C}{2}(kq)^2\left(B_c\cos qx-C\sin kx\right),\\
\partial^2a_1=-\frac{B_sC}{4}(kq)^2\left(
-{B}_c\left\{\cos(2q-k)x+\cos(2q+k)x\right\}+{C}\left\{\cos(2k-q)x+\cos(2k+q)x\right\}\right),\\
\sin(qx)\partial^2f_2-2\cos(qx)q_\alpha\partial^\alpha f_2=
-\frac{B_s^2}{2}(kq)^2\cos qx,\\
f_3=0,\\
\partial^2a_2=\frac{B_s^2C}{4}(kq)^2\left\{\cos(2k-q)x+\cos(2k+q)x\right\},
\end{eqnarray}
where $B_s=B\sin\Theta$, $B_c=B\cos\Theta$.

These equations are already free of foregoing pathologies and can be easily integrated. Finally, in the first order of approximation the vector potential of the system of two waves in space with spin noncommutativity reads
\begin{eqnarray}
\label{w1}
A^2=B_s\cos\left(kx+\frac{e^2}{4}\theta^4C^2(kq)(qx)+e^2\theta^4\varphi\right)+e^2\theta^4a_1,\\
\label{w2}
A^3=C\cos\left(qx+\frac{e^2}{4}\theta^4B_s^2(kq)(kx)\right)+B_c\cos(kx)+e^2\theta^4a_2.
\end{eqnarray}
where 
\begin{eqnarray*}
\varphi=\frac{B_cC}{4}(kq)\sin(q+k)x,\\
	a_1=\frac{B_sC}{16}(kq)\left(
	{B}_c\left\{\cos(2q-k)x-\cos(2q+k)x\right\}-{C}\left\{\cos(2k-q)x-\cos(2k+q)x\right\}\right),\\
	a_2=-\frac{B_s^2C}{16}(kq)\left\{\cos(2k-q)x-\cos(2k+q)x\right\}.
\end{eqnarray*}
The last expression shows that spin noncommutativity effectively changes the wave vector of some components of waves and produces higher harmonics. The latter effect is expectable because of nonlinearity of the electrodynamics.

From (\ref{w1}), (\ref{w2}) it also follows that two waves do not interact with each other if $\Theta=0$ or $\Theta=\pi$. This corresponds to the case where only one component of field is present (\ref{s3a}), (\ref{s3b}) and from assertions of Section 3 it should be so. The influence of spin noncommutativity is zero in the case of $(kq)=|\bm{k}||\bm{q}|-(\bm{k},\bm{q})=|\bm{k}||\bm{q}|(1-\cos\vartheta)=0$, where $\vartheta$ is the angle between $k^\mu$ and $q^\mu$. This corresponds to the propagation of waves in the same direction ($\vartheta=0$).

\subsection{Exact solution for a plane wave propagation in constant electric and magnetic fields}

Propagation of a plane wave in a constant magnetic field in space with canonical noncommutativity was studied in \cite{guralnik2001testing}. In this paper a modified dispersion relation for an electromagnetic wave was obtained in the first order of approximation in $\theta$. It reads
\begin{eqnarray}
\label{pol}
\omega=|\bm{k}|(1-\bm{\theta}_T\bm{b}_T),
\end{eqnarray}
where $\theta_i=\varepsilon_{ijk}\theta^{ij}/2$, and $T$ in index denotes the transverse to $\bm{k}$ component.

Problem of propagation of a plane wave in a constant magnetic and electric fields in space with spin noncommutativity can be solved exactly. 
Let us choose the vector potential in the following form
\begin{eqnarray}
\label{A00}
A_0^\mu=A_f^\mu+A_w^\mu,
\end{eqnarray}
where $A_f^\mu=\left(-\bm{E}\bm{x};\bm{a}\bm{x};\bm{b}\bm{x};0\right)^T$ is the potential of a constant electric and magnetic fields, the term $A_w^\mu=\left(0;0;0; A\cos(kx)\right)^T$ corresponds to the plane wave.
Potential $A_f^\mu$ in commutative space corresponds to the magnetic field $\bm{B}=(-b_z;a_z;b_x-a_y)^T$ and electric field $\bm{E}$. Note that in commutative space some components of $\bm{a},\;\bm{b}$ can be putted equal to zero by the corresponding gauge transformations, so the values of the external fields do not depend on these components. But in noncommutative space it cannot be done. Therefore, different sets of $\bm{a},\;\bm{b}$, which in commutative space correspond to the same field, correspond to different fields in noncommutative space. Vector potential with additional linear combination of coordinates in $A^3_f$ in commutative space also describes the constant magnetic field, but in noncommutative space such a problem cannot be solved exactly.

The first step in obtaining the solution is an observation that the first order corrections $A_{(1)}^0$, $A_{(1)}^1$, $A_{(1)}^2$ equal zero. It can be obtained from (\ref{eqel}) that
$$\partial^2A_{(1)}^0=0,\;\partial^2A_{(1)}^1=0,\;\partial^2A_{(1)}^2=0.$$
The equation for $A_{(1)}^3$ is a bit more complicated and reads
$$\partial^2A_{(1)}^3=\tilde{A}\cos(kx),$$
where $\tilde{A}=A\left[(\bm{E},\bm{k})^2-(\bm{a},\bm{k})^2-(\bm{b},\bm{k})^2\right]$. This equation is resonant. But consideration within the Bogolyubov-Krylov method as in the previous subsection shows that the solution of this equation is a plane wave with modified wave vector and what is important no nonlinearities appear in this case.

Equations for all the next terms of expansions of $A^\mu$ in $\theta^4$ are qualitatively the same: noncommutativity does not influence the potentials of a constant fields $A_{(n)}^0$, $A_{(n)}^1$, $A_{(n)}^2$ and changes the wave vector of the wave.

Using such an analysis within the perturbative scheme, we can conclude that the exact solution of the field equations (\ref{eqel}) is again a plane wave $A^3=A\cos(k_sx)$ which proliferates in a constant fields 
\begin{eqnarray}
\label{field_lin}
A^0=-\bm{Ex},\; A^1=\bm{ax},\; A^2=\bm{bx}.
\end{eqnarray}

Substitution of this ansatz into (\ref{eqel}) gives the following equations
\begin{eqnarray}
\label{sol2a}
\partial^2A^0=0;\quad\partial^2A^1=0;\quad\partial^2A^2=0;\\
\label{sol2b}
\partial^2A^3=-e^2\theta^4(E^2-a^2-b^2)\tilde{\partial}^2A^3-e^2\theta^4\left[(\bm{E}\tilde{\nabla})^2-(\bm{a}\tilde{\nabla})^2-(\bm{b}\tilde{\nabla})^2\right]A^3.
\end{eqnarray}	
Equations (\ref{sol2a}) are exact and are in complete agreement with the assertion (\ref{field_lin}) of linearity of $A^0$, $A^1$, $A^2$. From (\ref{sol2b}) it can be found an exact modified dispersion relation for $k_s^\mu$, which reads
\begin{eqnarray}
\label{disp0}
\omega^2_s=\bm{k}_s^2\frac{1+e^2\theta^4\left([{\bm{E},\bm{k}_s}]^2-[{\bm{a},\bm{k}_s}]^2-[{\bm{b},\bm{k}_s}]^2\right){\rm sinc}^2{(\theta k_s)}}{1+e^2\theta^4\left(E^2-a^2-b^2\right){\rm sinc}^2{(\theta k_s)}},
\end{eqnarray}
where $k_s=\sqrt{\omega_s^2-\bm{k}_s^2}$. This expression can be also rewritten in the following manner
\begin{eqnarray}
\label{disp}
k^2_s=-\frac{e^2\theta^4\left(({\bm{E},\bm{k}_s})^2-({\bm{a},\bm{k}_s})^2-({\bm{b},\bm{k}_s})^2\right){\rm sinc}^2{(\theta k_s)}}{1+e^2\theta^4\left(E^2-a^2-b^2\right){\rm sinc}^2{(\theta k_s)}},
\end{eqnarray}
Equation (\ref{disp}) has a unique solution for $k_s^2$ for a given set of $\bm{E},\;\bm{a},\;\bm{b}\text{ and }\bm{k}$. Therefore, there are no birefringence of a plane electromagnetic wave in external constant fields. The sign of  $k_s^2$ (sign of a squared mass $m^2=k_s^2$) depends on the sign of numerator in (\ref{disp}). In particular, in the pure electric field $k_s^2<0$ (since square of a mass is negative such a photons can be called like `tachionic' photons), in the magnetic field $k_s^2>0$ (photons with mass $m^2=k_s^2$).

It is interesting to compare obtained result with (\ref{pol}). In the first order of $\theta^4$ in the absence of electric field $\bm{E}$ (\ref{disp0}) reads
\begin{eqnarray}
\label{appp}
\omega=|\bm{k}|[1+e^2\theta^4(({\bm{a},\bm{e}_k})^2+({\bm{b},\bm{e}_k})^2)],
\end{eqnarray}
where $\bm{e}_k=\bm{k}/k$.
As we can see, in a space with spin noncommutativity the influence of a magnetic field on the wave dispersion relation is qualitatively different from that in the case of canonical noncommutativity. Spin noncommutativity always increases the multiplier near $|\bm{k}|$ in dispersion relation written in the manner of (\ref{appp}). Moreover, on the contrary to the case of canonical noncommutativity, in spin noncommutativity modifications of the dispersion relation depends on longitudinal with respect to $\bm{k}$ components of a magnetic field.
\section{Conclusions}
In the present paper we have introduced new Lorentz-invariant noncommutativity (\ref{alg}). Noncommutative position operators are built from the commutative coordinates by shifting the last by the corresponding Dirac gamma matrices $X^\mu=x^\mu+i\theta\gamma^\mu$. We have found the analogue of the Moyal star product in this space. Properties of such a product are quite different from the properties of the analogous product in canonical noncommutativity. We have found that every commutative function $f(x)$ has its noncommutative counterpart $\widehat{T}f(x)$ in space with spin noncommutativity. This counterpart is built by acting the translation operator $\widehat{T}=e^{i\theta(\gamma_\mu\partial^\mu)}$ on the commutative function. Also we have considered a product of noncommutative functions. This product is noncommutative, but it is associative. In Section 2 we have studied mathematical properties of such noncommutative functions in detail. 

In Section 3 an action for the electromagnetic field in space with spin noncommutativity has been built (\ref{act}), (\ref{act_2}). We have found the gauge transformation law of such a field (\ref{gauge}). These transformations resemble the gauge transformation of the Yang-Mills field. It is interesting that in the case of the abelian group $U(1)$ of electromagnetic field, the group of noncommutative electromagnetic field can be written as a direct product $U(1)\times SL(1,3)$.  Also the exact nonlinear field equations (\ref{eqel}) have been obtained from the least action principle. In addition, the structure of these equations has been analysed. Although these equations have been obtained for free field, interaction with an external current $\jmath^\mu$ can be easily included in this scheme (\ref{eqel_int}).

Nonlinearity of the field equations leads to the interaction between electric and magnetic fields. Within the framework of obtained electrodynamics we have considered the influence of a magnetic field on an electrostatic field of a point charge.  We have found that the magnetic field shields the charge in the Coulomb law, moreover such a screening is anisotropic and always decreases the effective charge value (\ref{sol1a}). Also we have studied an interaction of two waves in the space with spin noncommutativity within the perturbation theory. Ordinary expansion of potential $A^\mu$ in $\theta^4$ leads to resonant equations, which is an artifact of perturbations. To avoid resonant term we have used well-known from theoretical mechanics the Bogolyubov-Krylov method of studying nonlinear oscillations. Interaction of two electromagnetic plane waves changes the wave vector of each wave and produces higher-mode oscillations (\ref{w1}), (\ref{w2}). An exact solution of a plane wave propagation in constant magnetic and electric fields has been found. It has been showed that external field modifies the dispersion relation of the wave (\ref{disp0}). 

\section*{Acknowledgments}
This work was partly supported by Project FF-30F (No. 0116U001539) from the
Ministry of Education and Science of Ukraine.

\bibliographystyle{unsrt}
\bibliography{EL}{}

\end{document}